\documentclass[12pt, onecolumn, draftclsnofoot, journal]{IEEEtran}
\usepackage[dvips]{graphicx}

\usepackage[T1]{fontenc}
\usepackage{cite}
\usepackage[cmex10]{amsmath}
\usepackage[cmintegrals]{newtxmath}
\usepackage{amssymb,enumerate}
\usepackage{flushend}

\usepackage{comment}

\newcommand{\Prb}{\mathsf{P}}\newcommand{\Exp}{\mathsf{E}}

\newcommand{\dd}{\mathrm{d}}\newcommand{\ee}{\mathrm{e}}
\newcommand{\R}{\mathbb{R}}
\newcommand{\N}{\mathbb{N}}

\newcommand{\SINR}{\mathsf{SINR}}

\newcommand{\bsym}[1]{\boldsymbol{#1}}

\newtheorem{proposition}{Proposition}
\newtheorem{theorem}{Theorem}
\newtheorem{corollary}{Corollary}

\newtheorem{remark}{Remark}

\begin{document}

\title{Time-based Handover Skipping \\in Cellular Networks: Spatially Stochastic \\Modeling and Analysis}

\author{Kiichi~Tokuyama, Tatsuaki~Kimura, and~Naoto~Miyoshi%
\thanks{K. Tokuyama and N. Miyoshi are with the Department of Mathematical and Computing Science, Tokyo Institute of Technology, Tokyo, Japan. E-mail: tokuyama.k.aa@m.titech.ac.jp, and miyoshi@is.titech.ac.jp.}%
\thanks{T. Kimura is with the Department of Information and Communications Technology, Graduate School of Engineering, Osaka University, Osaka, Japan. E-mail: kimura@comm.eng.osaka-u.ac.jp.}%
\thanks{The support of the Japan Society for the Promotion of Science (JSPS) Grant-in-Aid for Scientific Research (C) 19K11838 is gratefully acknowledged.}}
\maketitle

\begin{abstract}

Handover (HO) management has attracted attention of research in the context of wireless cellular communication networks. One crucial problem of HO management is to deal with increasing HOs experienced by a mobile user. To address this problem, HO skipping techniques have been studied in recent years. In this paper, we propose a novel HO skipping scheme, namely, \textit{time-based HO skipping}\footnote{This work has been presented in part in \cite{TokuyamaMiyoshi18}.}. In the proposed scheme, HOs of a user are controlled by a certain fixed period of time, which we call \textit{skipping time}. The skipping time can be managed as a system parameter, thereby enabling flexible operation of HO skipping. We analyze the transmission performance of the proposed scheme on the basis of a stochastic geometry approach. In the scenario where a user performs the time-based HO skipping, we derive the analytical expressions for two performance metrics: the HO rate and the expected data rate. The analysis results demonstrate that the scenario with the time-based HO skipping outperforms the scenario without HO skipping particularly when the user moves fast. Furthermore, we reveal that there is a unique optimal skipping time maximizing the transmission performance, which we obtain approximately.
%Furthermore, we demonstrate that there is a unique optimal skipping time that maximizes the transmission performance, and derive an approximate expression of the optimal skipping time.

\end{abstract}


\section{Introduction}\label{sec:introduction}
\subsection{Background and Motivation}
The development of fifth generation (5G) wireless cellular communication systems is driven by
the need to satisfy the ever-increasing capacity demand resulting from the proliferation of mobile phones, tablets, and other handheld mobile devices. One of the key features of the 5G evolution is network densification through small cell deployment \cite{RomBitImr15}. Densifying base stations (BSs) shrinks the service area of each BS, increases spectral efficiency, and offers more capacity, thereby enabling a significant increase in the quality of service.
However, deploying more BSs promotes frequent handovers (HOs), which increase the risk of disconnection and signaling overhead. Therefore, countermeasures against the problem of frequent HOs are under discussion upon conducting network densification. 

The problem of frequent HOs has been addressed by means of HO skipping \cite{Arshad1, Arshad2, Arshad3, Arshad4, Demarchou, RodJayaDutk19, WuHaas19}, which is a technique whereby some HOs of a moving user are skipped to reduce excessive HOs. This approach is particularly effective when the user moves fast, or the network consists of densified small cells. The HO skipping enables a user to reduce the HO rate, whereas it may decrease the data reception rate (data rate) because it demands the user to maintain longer connection duration with a BS. Thus, the HO skipping presents a trade-off relation between the increased HO rate and the decreased data rate, and the trade-off relation should be balanced to improve transmission performance of users. Accordingly, a reliable framework for analyzing the trade-off relation is needed to evaluate the HO skipping. 
In this paper, we propose a novel HO skipping scheme, namely, {\it time-based HO skipping}, and provide an analysis framework for the two metrics, the HO rate and the expected data rate.

\subsection{Related Work}
\subsubsection{HO Management}
To suppress frequent HOs, HO management for mobile users has been considered. This management has recently attracted attention in the context of cellular networks (see, e.g., \cite{Tayyab19} for a survey). Zahran \textit{et al.}~\cite{ZahLiangSaleh06} proposed an application-specific signal threshold adaptation scheme to reduce excessive HOs. In the scheme, they established an algorithm for HO management by considering the received signal strength and estimated life time. Zhang \textit{et al.}~\cite{ZhangMaLi11} proposed an algorithm for reducing HO signaling overhead. However, BS intensity and interference signals from other BSs are not considered in their analysis. In addition, none of these studies describe the trade-off relation between increased HO rate and decreased data rate. Ylianttila \textit{et al.}~\cite{Ylianttila01} proposed dwell timer schemes for HO management, and they considered the trade-off relation to find the optimal HO frequency that improves the transmission performance. However, their study is based on simulations, and an analytical framework for evaluating the trade-off is not considered.

\subsubsection{HO Rate and Data Rate Analysis}
To investigate the effect of HOs on the transmission performance, the HO rate and the data rate are essential metrics. Motivated to provide mathematical studies for these metrics, theory of spatial point processes and stochastic geometry has been adopted to model cellular networks.
Baccelli and Zuyev~\cite{BaccZuye97} first analyzed the HO rate by a stochastic geometry approach. By modeling the BS locations in a cellular network with a homogeneous Poisson point process~(PPP), they derived analytical expressions for the distribution of the number of HOs during a fixed period of time when a user moves along a straight line.
Lin \textit{et al.}~\cite{LinGantFlemAndr13} proposed a modified
random waypoint~(RWP) model to describe user mobility.
By applying this model to homogeneous networks where BSs are arranged
according to a hexagonal grid or a homogeneous PPP, they derived
analytical results of the HO rate and the expected
sojourn time in a typical BS cell.
However, the above results do not consider the data rate.

Regarding data rate analysis by a stochastic geometry approach, Andrews \textit{et al.}~\cite{AndrBaccGant11} first analyzed the data rate and coverage focusing on signal-to-interference-plus-noise ratio (SINR) distribution in the context of cellular networks. They derived the analytical expressions of the coverage probability and the expected downlink data rate of a typical user in a single-tier homogeneous cellular network. Renzo \textit{et al.}~\cite{DiReGuidCora13} analyzed the downlink data rate of a typical user under generalized assumptions on the fading effect. However, the above studies assume that the typical user is static, and therefore they did not consider HOs. Analyzing the trade-off relation between increased HO rate and decreased data rate are crucial for evaluating the transmission performance of moving users. 

Several works studied the trade-off through analyzing the HO rate and the data rate.
Bao and Liang~\cite{BaoLian15} considered a multi-tier heterogeneous
network configured by overlaid independent homogeneous PPPs and
derived the exact expressions of the intra- and inter-tier HO
rates for a user moving on an arbitrary trajectory.
In addition, they proposed an optimal tier selection by balancing the HO rate and the expected downlink data rate.
Another study of Bao and Liang \cite{BaoLian16} considered BS cooperated cellular networks, and they provided analytical expressions for the HO and expected downlink data rates. In addition, they proposed the optimal cooperating strategy by maximizing an evaluation function consisting of those two metrics.
Chattopadhyay \textit{et al.}~\cite{ChatBlasAltm19} considered maximizing throughput in a two-tier network, where multiple moving users and static users are connected to either of macro or micro BSs. They found the optimal proportion of macro and micro BSs density and their respective transmit power levels by balancing the HO rate of the moving users, the expected downlink data rate of the moving and static users, and the expected number of the users connected to a BS, any of which are analytically described.

Some related works considered a trade-off relation between HO rate and coverage. Sadr and Adve~\cite{SadrAdve15} considered the HO rate in multi-tier heterogeneous cellular networks and analyzed the negative effect of performing HOs on the coverage. Using the analysis results, they derived the optimal proportion of the BS densities in different tiers to maximize the coverage.

\subsubsection{Analysis of HO Skipping}

HO skipping is an approach to reduce excessive HOs, performing HOs only when a certain predefined condition is satisfied and skipping HOs otherwise. By doing so, it aims to balance the relation between HO rate and data rate. 
Due to the effectiveness, many HO skipping schemes have recently been proposed, and some of the works successfully provided analytical frameworks for HO skipping schemes via stochastic geometry. 
A fundamental HO skipping scheme, which we call \textit{alternate HO skipping}, was introduced by Arshad \textit{et al.}~\cite{Arshad1}. In this scheme, a user alternately performs HOs along its trajectory, thereby achieving a 50\% reduction in the HO rate. Moreover, they analyzed the negative effect of the alternate HO skipping on the downlink data rate, that is, the impact of the prolonged connection duration with a BS due to the alternate HO skipping. Using these results, they evaluated user throughput by constructing an evaluation function, which represents the trade-off between increase in HO rate and decrease in data rate. It was demonstrated that the user throughput can be improved at least when either the moving speed of a user or the BS density is sufficiently high. The work \cite{Arshad1} was extended in \cite{Arshad2} and \cite{Arshad3}; \cite{Arshad2} introduced the alternate HO skipping in two-tier networks and \cite{Arshad3} introduced it in a BS cooperating network with coordinated multi-point (CoMP) transmission.
Another HO skipping scheme, called \textit{topology-aware HO skipping}, was proposed in \cite{Arshad4}. In this scheme, the HO skipping is triggered according to the user's distance from the target BS and the size of the cell; that is, HOs are not performed when the user enters a cell whose area is smaller than a certain threshold, or a cell whose dominant BS is farther from the user than a certain threshold. 
%This scheme is particularly effective for suppressing prolonged connection distance caused by HO skipping. 
%The topology-aware HO skipping is particularly effective for suppressing prolonged connection distance caused by HO skipping, since positional aspects are primarily considered in this scheme. 
Thus, the topology-aware HO skipping can prevent the connection distances from being unnecessarily long. However, this scheme was evaluated by Monte Carlo simulations, and no mathematical framework for the performance evaluation was presented. The mathematical framework for evaluating topology-aware HO skipping was provided by Demarchou \textit{et al.}~\cite{Demarchou}. They derived the analytical expressions of the coverage probability and the expected downlink data rate when the topology-aware HO skipping is performed under the assumption that the trajectory of a moving user is a straight line. They also proposed an HO skipping scheme that is suited to a user-centric CoMP scheme. 
However, both the alternate and the topology-aware HO skippings do not consider the connection duration, i.e., the communication time connecting to a BS until the next HO. 
Such connection duration is essential for the HO management due to the problem of signaling overhead. More precisely, when a user performs an HO, its data transmission is temporarily disrupted during the signaling procedure of the HO. This sudden disruption causes unwanted delay and may significantly degrade the performance of network applications. Thus, keeping connection for a certain duration is important for the communication quality of ultra-reliable applications, e.g., high-quality video streaming services and vehicular safety applications. Owing to the above reason, an HO skipping scheme that can control the connection duration needs developing.
%Such connection duration should be managed for the HO management regarding the problem of signaling overhead. Since signaling to the target BS when performing an HO causes a temporary interruption of transmitting data, connections seriously relying on the data transmission at each moment might suffer from a sudden delay caused by irregularly performed HOs.
%Therefore, ensuring a certain extent of time of a connection, managing the connection duration is required for the quality of service of ultra-reliable mobile applications, such as video streaming services and vehicular safety applications. From the above reason, an HO skipping scheme that can control the connection duration needs developing.

\subsection{Contribution of This Study}
In this paper, we propose a novel HO skipping scheme, \textit{time-based HO skipping}. Our proposed scheme considers a threshold of time, which we call the \textit{skipping time}, to control the frequency of HOs for a moving user; that is, all HOs attempted earlier than the threshold are skipped, and a user is guaranteed to keep connected during this threshold time.
By adjusting the skipping time, our proposed scheme enables to directly control the connection duration of the user with a BS. 
%Our proposed scheme provides flexible operation of HO skipping in the sense that adjusting the threshold of time, which we call the \textit{skipping time}, enables to control how long a user skips HOs.
We provide a tractable framework for analyzing the time-based HO skipping scheme. Under a random walk model of user mobility, we derive the analytical expressions of the HO rate and the expected downlink data rate when the user performs the time-based HO skipping. Moreover, by using these two metrics, we construct an evaluation function of the transmission performance of a moving user, regarding the trade-off relation between the HO rate and the data rate. On the basis of this evaluation function, we conduct a performance comparison between two scenarios: (1) the non-skipping scenario where the moving user do not perform any HO skipping and experience HOs whenever they cross the boundaries of BS cells, and (2) the HO skipping scenario where the moving user perform the time-based HO skipping. The results indicate that the difference of the performance in these two scenarios depends on the speed of the moving user; the non-skipping scenario is better when the user moves slow, whereas the HO skipping scenario is better when the user moves fast.
In addition, we maximize the evaluation function by controlling the skipping time of the time-based HO skipping scheme. We also investigate the relation among the optimal skipping time that maximizes the evaluation function and the other system parameters in our model. Furthermore, we attempt to derive an approximate expression of the optimal skipping time.
This paper gives much enhancement of \cite{TokuyamaMiyoshi18}; enriched analysis results are provided compared with \cite{TokuyamaMiyoshi18}, and the optimal skipping time is newly found in this paper. 
All the results of this paper are verified by numerical experiments.

\subsection{Paper Organization}
The rest of the paper is organized as follows. In Section~\ref{sec:model}, we describe the network and user mobility models, propose the time-based HO skipping scheme, and define the evaluation function of the transmission performance. The analyses of the HO rate and the data rate are provided in Section~\ref{sec:analysis}. In Section~\ref{sec:evaluation}, we evaluate the transmission performance with numerical experiments, and consider maximizing the transmission performance with respect to the skipping time in numerical and analytical ways. Finally, the paper is concluded in Section~\ref{sec:conclusion}.

\section{System Model}\label{sec:model}
\subsection{Network Model}

We consider a homogeneous cellular network where all BSs transmit
signals with the same power level (normalized to 1) utilizing
a common spectrum bandwidth.
We adopt the conventional assumption that the BSs are arranged according
to a homogeneous PPP~$\Phi$ on $\R^2$ with
intensity~$\lambda$~($\in(0,\infty)$), where the points~$X_1,
X_2,\ldots$ of $\Phi$ are numbered in an arbitrary order, and we assume that a user equipment (UE) receives downlink signals from one of the allocated BSs.
We also assume Rayleigh fading and power-law path-loss on the
downlinks but ignore shadowing effects; that is, when a UE at a
position~$\bsym{u}\in\R^2$ receives a signal from the BS located
at~$X_i\in\Phi$, $i\in\N:=\{1,2,\ldots\}$, at time~$t$ ($\in\N_0:=\N\cup\{0\}$), the received signal power is represented
by $H_{i,t}\|X_i-\bsym{u}\|^{-\beta}$, where $H_{i,t}$,
are mutually
independent and exponentially distributed random variables with unit
mean~($H_{i,t}\sim\mathrm{Exp}(1)$) representing the fading effects,
and $\beta$~($>2$) denotes the path-loss exponent. Here, $\| \cdot \|$ denotes the Euclidean norm.
%Note that $\|\cdot\|$ is Euclidean distance.

We suppose that each UE is initially associated with the
nearest BS; that is, the BS cells form a Poisson--Voronoi tessellation.
Owing to the stationarity of the network model, there is no loss of
generality in focusing on a UE that is assumed to be at the origin at
time zero; we refer to this UE as the typical UE.
Let $b_o$ denote the index of the nearest point of $\Phi$ to the
origin; that is, $\| X_{b_o} \| \le \| X_j \|$ for all  $j \in \N$.
Then, the probability density function of $R_{b_o}=\|X_{b_o}\|$ is given as follows, (see, e.g., Sec.~2.3 of~\cite{ChiuStoyKendMeck13})
\begin{equation}\label{eq:nearest_dens}
  f_{b_o}(r)
  = 2\pi\lambda\, r\,\ee^{-\lambda\pi r^2},
  \quad r\ge0.
\end{equation}

We further suppose that the UEs move on the two-dimensional plane,
and a BS transmits a signal to each of its serving UEs at every
unit of time.
When the typical UE is at a position~$\bsym{u}\in\R^2$ at
time~$t\in\N_0$ and is associated with the BS at $X_i\in\Phi$, the
downlink SINR of this UE is
represented by
\begin{equation}\label{eq:SINR}
  \SINR_{u,i,t}
  = \frac{H_{i,t}\|X_i-\bsym{u}\|^{-\beta}}
         {\sigma^2 + I_{u,i,t}},
\end{equation}
where $\sigma^2$ denotes a positive constant representing the noise
power, and $I_{u,i,t}$ denotes the interference power to the typical UE
given by
\begin{equation}\label{eq:interference}
  I_{u,i,t}
  = \sum_{j\in\N\setminus\{i\}}
      \frac{H_{j,t}}{\|X_j-\bsym{u}\|^\beta}.
\end{equation}
Then, $\tau_{u,i,t}$, the expected downlink data rate when the UE is associated with $X_i$ at a position $\bsym{u}$ at time $t$, is defined as
\begin{equation}\label{eq:datarate}
  \tau_{u,i,t}(\lambda,\beta)
  = \Exp[\log(1+\SINR_{u,i,t})].
\end{equation}

\subsection{Mobility Model with Time-based HO Skipping}\label{subsec:mobility}
\begin{figure}[!t]
\begin{center}
%\centering
\includegraphics[width =.47\linewidth, clip]{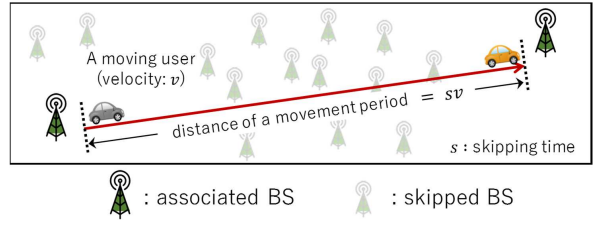}
\vspace{-1.30mm}
\caption{Illustration of the typical UE moving for a movement period in the time-based HO skipping model.
\label{fig:fig2}
}
\end{center}
\vspace{-3.80mm}
\end{figure}

We describe the time-based HO skipping scheme performed in a mobility model for a moving UE.
%We suggest a simple and tractable random walk model to describe the
%mobility of a UE, and explain the time-based HO skipping scheme performed in the random walk model.
First, we introduce a parameter $s$ as the {\it skipping time} of the typical UE and assume that the typical UE skips to perform HOs for the skipping time $s$; that is, the typical UE is allowed to perform an HO every $s$ units of time only.
We refer to each period of the skipping time $s$ as a {\it movement period}.
Fig.~\ref{fig:fig2} illustrates the typical UE moving straightly for a movement period. We note that the UE is not always associated with the closest BS during the movement period, and the expected distance between the UE and its associated BS becomes larger as the skipping time $s$ becomes larger. Whereas, the UE is associated with the closest BS in the end of the movement period since the UE can perform HOs at the endpoint.

\begin{figure}[!t]
\begin{center}
\includegraphics[clip,width=.45\linewidth]{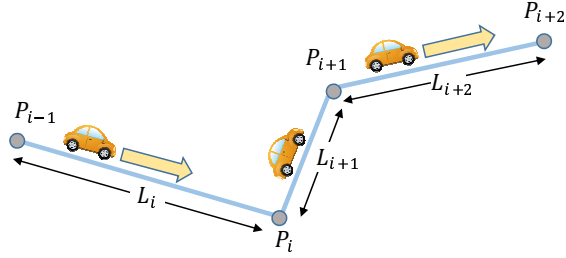}
\end{center}
\caption{Illustration of a trajectory of the typical UE in the random walk mobility model.}
\label{fig:trajectory}
\end{figure}

We next give a simple and tractable random walk model to describe mobility of the typical UE.
We recall that the typical UE is at the origin at time zero. Let $\{Y_1,Y_2,\ldots\}$ denote a sequence of independent and
identically distributed (i.i.d.) random variables on $\R^2$.
Then, the position of the typical UE after $n$~movement periods is
given by the two-dimensional random walk
\[
  P_n = \sum_{k=1}^n Y_k, \quad n=1,2,\ldots,
\]
with $P_0=o=(0,0)$.
We assume that the typical UE moves on the line segment at a constant velocity during each movement period; that is, the velocity $V_n$ during the $n$-th
movement period is equal to $Y_n/s$ for $n=1,2,\ldots$.
A trajectory of the typical UE is shown in
Fig.~\ref{fig:trajectory}.
This random walk model is a flexible mobility model that can capture various mobility patterns by choosing the distribution of $Y_n$.
Let $(L_n, \psi_n)$ denote the polar coordinates of $Y_n$.
The distributions of $L_n$ and $\psi_n$ respectively represent the variation of a moving distance, and a direction change of the UE in the $n$-th movement period. Note that the distance $L_n$ corresponds to the moving speed of the UE in the $n$-th movement period since we consider the common moving time $s$ in every movement period.
For instance, $\Prb(L_n=0)>0$ represents that the UE can take a pause in $s$~units of time.
Furthermore, if $\psi_n$ is a constant, the UE always moves in a straight line.

\begin{figure}[!t]
\begin{center}
%\centering
\includegraphics[width =.4\linewidth, clip]{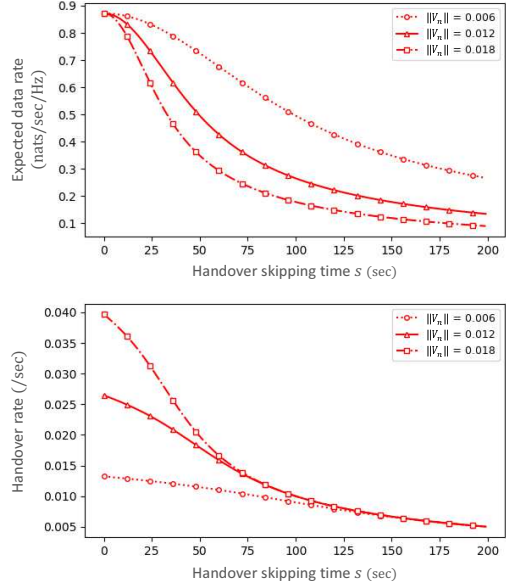}
\vspace{-1.30mm}
\caption{Numerical example of the expected data rate (upper) and HO rate (lower) with three cases of the moving speed of users, where the other parameters are set as $\lambda=3$ (units/km$^2$) and $\beta=3$. This result is obtained by the analysis developed in the following section.
\label{fig:fig3}
}
\end{center}
\vspace{-3.80mm}
\end{figure}

We consider the expected data rate and the HO rate of the typical UE in this model of the time-based HO skipping scheme. Fig.~\ref{fig:fig3} shows numerical examples of the expected data rate and the HO rate of the UE attempting the time-based HO skipping with skipping time $s$, which are based on the analysis in the next section. In this example, there are three parameter patterns as follows: $\|V_n\|:=L_{n}/s=0.006, 0.012, 0.018$ (km/s), which denotes the speed of the UE moving in the $n$-th movement period. It can be seen that the expected data rate decreases as the skipping time $s$ becomes large. 
This is because the UE moves to a further position from the origin as $s$ becomes larger, so that the expected data rate becomes smaller when the UE fixes its associated BS.
On the other hand, it can also be seen that the HO rate decreases as $s$ increases. Intuitivelly, this is because at most one HO occurs during a movement period of time $s$ so that the HO rate, that is the expected number of performed HOs divided by $s$, decreases as $s$ increases.

To conduct comparison studies to evaluate the time-based HO skipping scheme, we consider two scenarios; Scenario~1 is the non-skipping scenario: the typical UE does not attempt HO skipping and performs HOs whenever the typical UE crosses the boundaries of cells. Scenario~2 is the time-based HO skipping scenario: the typical UE attempts the time-based HO skipping in every movement period.

\subsection{Evaluation Function of Transmission Performance}
We define an evaluation function of transmission performance of a moving UE, so that the evaluation function represents the trade-off between increase in HO rate and decrease in data rate.
%increased HO rate and decreased data rate. 
Let $D_{m}(s, l, \lambda, \beta)$ and $\mathcal{N}_{m}(l, \lambda)$, $m=1,2$, denote the expected amounts of received data and the expected number of HOs for the typical UE moving through a distance $l$ for time $s$ in Scenario~$m$, respectively.  
We note that $s$ corresponds to the skipping time, and $l$ corresponds to a moving distance during a movement period. 
Using these expressions, we define the transmission performance in the respective scenarios as follows:
\begin{align}
\label{eq:tpfunc}
&\!\!\!Q_{m}(s, \lambda, \beta, \mathcal{C})
\nonumber\\
&=\frac{1}{s}\big\{\Exp[D_{m}(s, L_{n}, \lambda, \beta)]-\mathcal{C}\Exp[\mathcal{N}_{m}(L_{n}, \lambda)]\big\},\quad m=1, 2,
\end{align}
where $\mathcal{C}$ is a cost factor that transforms an HO into the amount of lost data by conducting the HO\footnote{As described in Section~\ref{sec:introduction}, there are some risks of performing HOs such as the disconnection and the signaling overhead. In our model, such risk factors of HOs are transformed to the amount of lost data by the constant $\mathcal{C}$.}.
We note that $\Exp[D_{m}(s, L_n, \lambda, \beta)]/s$, $m=1,2$, denotes the data rate, whereas $\Exp[\mathcal{N}_{m}(L_n, \lambda)]/s$ denotes the HO rate for Scenario~$m$.
We also note that $\Exp[\mathcal{N}_{2}(L_n, \lambda)]$ is not greater than one, namely, the HO rate in Scenario~2 is not greater than $1/s$. This is due to the assumption that at most one HO occurs during the period of time $s$ in Scenario~2. Therefore, the term that has a negative impact on the performance of HOs are almost negligible in Scenario~2 when $s$ is sufficiently large.

\section{Handover Rate and Data Rate Analysis}\label{sec:analysis}

In this section, we analyze the transmission performance of a moving UE.
We investigate the HO rate and the expected downlink data rate
for the typical UE in Scenarios~1 and~2 described in the preceding section.
We note that in Scenario~2, if the typical UE has crossed one or more boundaries
of BS cells in a movement period, then it experiences an HO at the end of the period and is associated with the nearest BS from the new position.
Throughout this section, we focus only on the first movement period
and fix the movement of the typical UE as $Y_1 = (l,0)$ for $l\ge0$;
that is, the typical UE starts from the origin and moves to $(l,0)$ in
$s$~units of time.

\subsection{Handover Rate Analysis}

By the definition, $\mathcal{N}_1(l,\lambda)$ is evaluated as the expected
number of intersections of a line segment of length~$l$ and
the boundaries of Poisson--Voronoi cells. On the other hand, $\mathcal{N}_2(l,\lambda)$ is evaluated as the probability that a line segment of length~$l$ has one or more
intersections with the boundaries of Poisson--Voronoi cells because at most one HO occurs in a movement period of a distance~$l$. As described in Remarks~\ref{rmk:handover1} and~\ref{rmk:handover2} below, the analytical expressions of $\mathcal{N}_1(l,\lambda)$ and $\mathcal{N}_2(l,\lambda)$ are both obtained from the existing results. For this reason, we leave the proofs to the references, and introduce only the results of these expressions of $\mathcal{N}_1(l,\lambda)$ and $\mathcal{N}_2(l,\lambda)$.

\subsubsection{Handover Rate Analysis for Scenario~1}

The HO rate in Scenario~1, that is the non-skipping scenario, is given as follows.
\begin{proposition}\label{prp:handover1}
The expected number of HOs $\mathcal{N}_1(l, \lambda)$ in the non-skipping scenario is given by
\begin{equation}\label{eq:handover1}
  \mathcal{N}_1(l, \lambda)
  = \frac{4\sqrt{\lambda}\,l}{\pi}.
\end{equation}
\end{proposition}

\begin{remark}\label{rmk:handover1}
This result was previously proved in various studies in the context of HO rate analysis related to Poisson--Voronoi cells. We are referred to e.g., \cite{BaccZuye97}, \cite{LinGantFlemAndr13} or \cite{BaoLian15} for the comprehensive proof.
\end{remark}

\subsubsection{Handover Rate Analysis for Scenario~2}

The HO rate in Scenario~2, that is the time-based HO skipping scenario, is given as follows.
\begin{proposition}\label{prp:handover2}
The expected number of HOs $\mathcal{N}_2(l, \lambda)$ in the time-based HO skipping scenario is given by
\begin{equation}\label{eq:handover2}
  \mathcal{N}_2(l, \lambda)
  =
 1 - 2\lambda
 	\int^{\pi}_{0}
 		\int^{\infty}_{0}
 			r\ee^{-\lambda\{\pi r^{2}+B(l, r, \theta)\}}
 		\dd r
	\dd \theta,
\end{equation}
where
\begin{equation}\label{eq:regionB}
  B(l, r, \theta)={w_{l, r, \theta}}^2\Big[\theta+\sin^{-1}\Big(\frac{l\sin\theta}{w_{l, r, \theta}}\Big)\Big]-r^2\theta+lr\sin\theta,
\end{equation}
with
\begin{equation}\label{eq:wcos}
  w_{l, r, \theta}=\sqrt{l^{2}+r^{2}-2lr\cos\theta}.
\end{equation}
\end{proposition}
\begin{remark}\label{rmk:handover2}
This result was proved by~\cite{SadrAdve15} because the definition of the HO rate in~\cite{SadrAdve15} corresponds to our HO rate in Scenario~2. We are referred to the proof of {\it Theorem~1} in ~\cite{SadrAdve15} for verification of the result in \eqref{eq:handover2}.
\end{remark}

\begin{figure}[!t]
\begin{center}
%\centering
\includegraphics[width =.4\linewidth, clip]{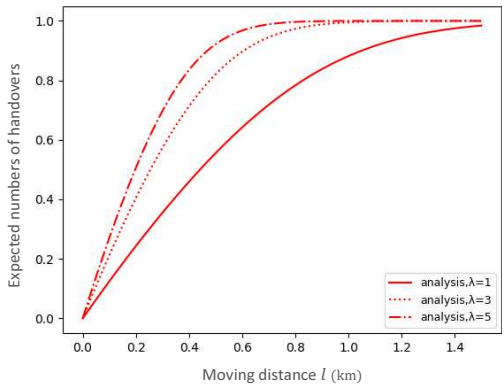}
\vspace{-1.30mm}
\caption{Numerical results of the analysis of $\mathcal{N}_2(l, \lambda)$. Three different cases of the parameter $\lambda$ are examined.
\label{fig:HO_compare}
}
\end{center}
\vspace{-3.80mm}
\end{figure}

Fig.~\ref{fig:HO_compare} shows numerically computed results of  $\mathcal{N}_2(l, \lambda)$ in \eqref{eq:handover2}. It can be seen that the expected number of HOs becomes larger when the intensity $\lambda$ of BSs becomes larger. This is because a UE tends to cross more cell boundaries as BSs are more densified. Besides, the expected number of HOs asymptotically approaches to $1$, since a UE surely crosses the boundary at least once when the UE moves through sufficiently long distance. 

\subsection{Data Rate Analysis}

We now consider $D_1(s, l, \lambda, \beta)$ and $D_2(s, l, \lambda, \beta)$, the expected amount of data received by the typical UE per movement period in Scenario~1 and Scenario~2, respectively.
We note that in Scenario~1, the typical UE is always associated to the nearest BS during a movement period, whereas the UE is not always associated to the nearest BS in Scenario~2.
We recall that the movement period is through a distance~$l$ for $s$~units of time.

\subsubsection{Data Rate Analysis for Scenario~1}

First, we provide an expression for $D_1(s, l, \lambda, \beta)$, which is
derived by applying Theorem~3 in~\cite{AndrBaccGant11}.

\begin{theorem}\label{prp:datarate1}
The expected amount of received data $D_1(s, l, \lambda, \beta)$ in the non-skipping scenario is given by
\begin{equation}\label{eq:datarate1}
  D_1(s, l, \lambda, \beta)
  = s\,\tau_1(\lambda, \beta),
\end{equation}
where
\begin{equation}\label{eq:tau1}
  \tau_1(\lambda, \beta)
  =\int_0^\infty\!\!\int_0^\infty
     \frac{\rho_1(z,u,\lambda,\beta)}{1+z}\,
   \dd z\,\dd u,
\end{equation}  
with
\begin{align*}
  \rho_1(z,u,\lambda,\beta)
  &= \exp\biggl\{
       -\sigma^2 z\,\biggl(\frac{u}{\pi\lambda}\biggr)^{\beta/2}
  \\
  &\quad\mbox{}
       - u\,\biggl(
           1 + \frac{2z^{2/\beta}}{\beta}
               \int_{1/z}^\infty
                 \frac{v^{2/\beta-1}}{1+v}\,
               \dd v
         \biggr)
     \biggr\}.  
\end{align*}
\end{theorem}
\begin{IEEEproof}
We assume that the typical UE is at a position~$\bsym{u}=(u,0)$, $0\le
u<l$, on the trajectory at time~$t$. We recall that $t$ ($=0, \cdots, s-1$) is defined as discrete time in our model, and note that $u = l t/s$ because the speed is constant in a movement period.
Let $b_u$ denote the index of the nearest BS from the
position~$\bsym{u}$, and let $R_u=\|X_{b_u}-\bsym{u}\|$.
Then, by \eqref{eq:SINR} and~\eqref{eq:datarate}, the data rate per unit time received at the position~$\bsym{u}$ is given by
\begin{equation}\label{eq:AmountData}
  \xi_1(u,\beta)
  = \log\biggl(
          1 + \frac{H_{b_u,t}\,R_u^{-\beta}}{\sigma^2 + I_{u,b_u}}
        \biggr),
\end{equation}
and we have
\begin{equation}\label{eq:expected_data1}
  D_1(s, l, \lambda, \beta)
  = \sum_{t=0}^{s-1}
      \Exp\Bigl[\xi_1\Bigl(\frac{l t}{s}, \beta\Bigr)\Bigr].
\end{equation}
Since the point process~$\Phi$ is stationary (the distribution is
invariant under translations), the distribution of $X_{b_u}-\bsym{u}$
is identical to that of $X_{b_o}$.
Furthermore, since $H_{i,t}$, $i\in\N$, $t\in\{1,2,\ldots,s\}$, are
i.i.d., $I_{u,b_u}$ and therefore $\xi_1(u, \beta)$ in
\eqref{eq:AmountData} are identically distributed for any $u$.
Thus, \eqref{eq:expected_data1} reduces to
\[
  D_1(s, l, \lambda, \beta)
  = s\,\Exp[\xi_1(0, \beta)].
\]
The expectation on the right-hand side above is the expected data
rate per unit of time considered in Theorem~3
of~\cite{AndrBaccGant11}.
Hence, by the conventional way (c.f. \cite{AndrBaccGant11}, \cite{Hamd10}), we obtain \eqref{eq:datarate1} with \eqref{eq:tau1}.
\end{IEEEproof}
\begin{remark}
Theorem~\ref{prp:datarate1} implies that when a moving UE is
always associated with its nearest BS, the expected amount of received data in $s$ units of time equals to $s$ times the expected data rate for a static UE, which is followed by the stationarity assumption.
A similar description is found in Remark~2
of~\cite{ChatBlasAltm19}. For this reason, $D_{1}(s, l, \lambda, \beta)$ does not depend on the moving distance $l$, that is, $D_{1}(s, l, \lambda, \beta)=D_{1}(s, \lambda, \beta)$. 
\end{remark}

\subsubsection{Data Rate Analysis for Scenario~2}

By contrast to Scenario~1, a different situation is observed in Scenario~2, where a UE
is not always associated with its nearest BS but remains
associated with the BS that is the nearest to the starting point.

\begin{theorem}\label{thm:datarate2}
The expected amount of received data $D_2(s, l, \lambda, \beta)$ in the time-based HO skipping scenario is given by
\begin{equation}\label{eq:D20}
  D_2(s, l, \lambda, \beta)
  = \sum_{t=0}^{s-1}
      \tau_2\bigl(\frac{l t}{s}, \lambda, \beta\bigr),
\end{equation}
where
\begin{align}
  &\tau_2(u, \lambda, \beta) 
  \nonumber\\
  &\!= \lambda\!
       \int_0^{2\pi}\!\!\!\!\int_0^\infty\!\!\!\!\int_0^\infty\!\!\!
         r\,\ee^{-\lambda\pi r^2-\sigma^2 w_{u,r,\theta}^{\,\beta}\, z}\,
           \frac{\rho_2(z,r,\theta|u,\lambda,\beta)}{1+z}\,
         \dd z
       \dd r\dd\theta,
  \label{eq:tau2}
\end{align}
with
\begin{align}
  &\!\!\!\!\!\!\!\!\!\!\!\!\!\!\!\!\!\!\!\!\!\!\rho_2(z,r,\theta|u,\lambda,\beta)
  \nonumber\\
  &\!\!\!\!\!\!\!\!\!\!\!\!\!\!\!\!\!\!\!\!\!\!\!= \exp\biggl(
        -\lambda
         \int_0^{2\pi}\!\!\!\!
          \int_r^{\infty}
            \Bigl\{
              1 + \frac{1}{z}\,
                  \Bigl(\frac{w_{u,v,\varphi}}{w_{u,r,\theta}}\Bigr)^\beta
            \Bigr\}^{-1}v\, 
          \dd v
         \dd\varphi
     \biggr),
  \nonumber
\end{align}
where $w_{u,v,\varphi}$ and $w_{u,r,\theta}$ are given in \eqref{eq:wcos}.
\end{theorem}
\begin{IEEEproof}
As in the proof of Theorem~\ref{prp:datarate1}, we suppose that
the typical UE is at a position~$\bsym{u}=(u,0)$, $0\le u<l$, on
the trajectory at time~$t$.
We recall that $X_{b_o}$ denotes the nearest point of $\Phi$ to the
origin.
Let~$(R_{b_o}, \Theta)$ denote the polar coordinate of $X_{b_o}$.
Then, the distance from the typical UE to $X_{b_o}$ is given by
\[
  R_{u,b_o}
  = \|X_{b_o} - \bsym{u}\|
  = \sqrt{
      {R_{b_o}\!}^2 + u^2 - 2R_{b_o} u\,\cos\Theta
    },
\]
where we should note that $R_{0,b_o} = R_{b_o}$.
An example image of the arrangement of the UE, $X_{b_o}$, and $X_i$ ($\in\Phi\setminus\{X_{b_o}\}$), is shown in
Fig.~\ref{fig:movingUE}.
\begin{figure}[!t]
\begin{center}
\includegraphics[clip,width=.45\linewidth]{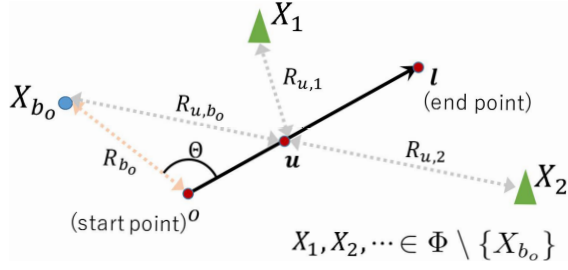}
\end{center}
\caption{Example image of arrangement of BSs and the typical
  UE.}
\label{fig:movingUE}
\end{figure}  
The data rate per unit time received at the position~$\bsym{u}$ is
given by
\begin{equation}\label{eq:xi20}
  \xi_2(u, \beta)
  = \log\biggl(
      1 + \frac{H_{b_o,t}\,R_{u,b_o}^{-\beta}}
               {\sigma^2 + I_{u,b_o}}
    \biggr),
\end{equation}
and we have
\begin{equation}\label{eq:D2}
  D_2(s, l, \lambda, \beta)
  = \sum_{t=0}^{s-1}
      \Exp\Bigl[
        \xi_2\bigl(\frac{l t}{s}, \beta\bigr)
      \Bigr].
\end{equation}
We note that the following equality holds from \eqref{eq:D20} and \eqref{eq:D2},
\begin{equation}\nonumber
\tau_2(u, \lambda, \beta)
=
\Exp\bigl[
  \xi_2(u, \beta)
\bigr].
\end{equation}
That is, $\tau_2(u, \lambda, \beta)$ in~\eqref{eq:tau2} corresponds to the expected data rate per unit of time for a UE at the
position~$\bsym{u}=(u,0), 0\leq u<l$ when the UE is associated with the nearest BS to the origin.
  
As $R_{b_o}\sim f_{b_o}$ in~\eqref{eq:nearest_dens} and $\Theta\sim
U(0,2\,\pi)$, we have 
\begin{align}\label{eq:xi2}
  &\Exp[\xi_2(u, \beta)]
   = \int_0^{2\pi}\!\!\!\int_0^\infty
       \lambda r\,\ee^{-\lambda \pi r^2}\,
  \nonumber\\
  &\qquad\mbox{}\times
       \Exp\biggl[
         \log\biggl(
           1 + \frac{H_{b_o,t}\,w_{u,r,\theta}^{-\beta}}
                    {\sigma^2 + I_{u,b_o}}
         \biggr)
       \biggm| R_{b_o}=r\biggr]\,
     \dd r\dd\theta,
\end{align}
where $w_{u,r,\theta}$ is given in~\eqref{eq:wcos}.
Note here that $H_{i,t}\sim\mathrm{Exp}(1)$ due to the assumption of Rayleigh fading. Then, applying Lemma~1 of~\cite{Hamd10} to the conditional expectation with respect to the event $\{R_{b_o}=r\}$ yields
\begin{align}\label{eq:cond_expect}
  &\Exp\biggl[
         \log\biggl(
           1 + \frac{H_{b_o,t}\,w_{u,r,\theta}^{-\beta}}
                    {\sigma^2 + I_{u,b_o}}
         \biggr)
       \biggm| R_{b_o}=r\biggr]
  \nonumber\\
  &=\int_0^\infty
       \frac{\exp(-\sigma^2\, w_{u,r,\theta}^{\,\beta}\, z)
         \bigl(1-\Exp\bigl[\ee^{-zH_{b_o,t}}\bigr]\bigr)}{z}
     \nonumber\\
     &\qquad\mbox{}\times   
     \Exp\bigl[
         \exp\bigl(
           - w_{u,r,\theta}^{\,\beta}\, z\,I_{u,b_o}
         \bigr)
       \bigm| R_{b_o}=r \bigr]\,
     \dd z,
  \nonumber\\
  &= \int_0^\infty
       \frac{\exp(-\sigma^2\, w_{u,r,\theta}^{\,\beta}\, z)}{1+z}
  \nonumber\\
  &\qquad\mbox{}\times     
       \Exp\bigl[
         \exp\bigl(
           - w_{u,r,\theta}^{\,\beta}\, z\,I_{u,b_o}
         \bigr)
       \bigm| R_{b_o}=r \bigr]\,
     \dd z,
\end{align}
where the Laplace transform of $H_{b_o,t}\sim\mathrm{Exp}(1)$ is applied in the last equality.
Furthermore, the interference expression in~\eqref{eq:interference}
leads to
\begin{align}
  &\Exp\bigl[
     \exp\bigl(
       - w_{u,r,\theta}^{\,\beta}\,z\,I_{u,b_o}
     \bigr)
   \bigm| R_{b_o}=r \bigr]
  \nonumber\\
  &= \Exp\biggl[
       \prod_{i\in\N\setminus\{b_o\}}
         \exp\Bigl\{
           - z\,H_{i,t}\,
             \Bigl(\frac{w_{u,r,\theta}}{R_{u,i}}\Bigr)^\beta
         \Bigr\}
     \biggm| R_{b_o}=r\biggr]
  \nonumber\\
  &= \Exp\biggl[
       \prod_{i\in\N\setminus\{b_o\}}
         \Bigl\{
           1 + z\,\Bigl(\frac{w_{u,r,\theta}}{R_{u,i}}\Bigr)^\beta
         \Bigr\}^{-1}
     \biggm| R_{b_o}=r\biggr]
  \nonumber\\
  &= \exp\biggl(
       -\lambda
         \int_0^{2\pi}\!\!\!\!
          \int_r^{\infty}
            \Bigl\{
              1 + \frac{1}{z}\,
                  \Bigl(\frac{w_{u,v,\varphi}}{w_{u,r,\theta}}\Bigr)^\beta
            \Bigr\}^{-1}v\, 
          \dd v
         \dd\varphi
     \biggr),
  \label{eq:cond_expect2}
\end{align}
where we use the Laplace transform of $H_{i,t}\sim\mathrm{Exp}(1)$,
$i\in\N\setminus\{b_o\}$, in the second equality, and the third equality follows from the Laplace functional of a homogeneous PPP (see, e.g., Chap.~4 in \cite{ChiuStoyKendMeck13}). We note that $w_{u,v,\varphi}$ in the third equality is given in \eqref{eq:wcos} as well as $w_{u,r,\theta}$.

Finally, substituting \eqref{eq:xi2}, \eqref{eq:cond_expect}, and \eqref{eq:cond_expect2} into
\eqref{eq:D2} leads to the result.
\end{IEEEproof}

We note that, in contrast to Scenario 1, $D_2(s, l, \lambda, \beta)$ in Theorem~\ref{thm:datarate2} is a function of $l$ because it depends on the speed of a UE. 

Theorem~\ref{thm:datarate2} provides the exact expression of the data rate in Scenario 2; however, the obtained expression is rather complex. If we are allowed to introduce an approximation, we can derive another simpler expression under the interference-limited assumption.  
\begin{corollary}\label{cor:datarate2}
If we assume $\sigma^2=0$, that is, noise power is negligible, then $\tau_2(u, \lambda, \beta)$ in \eqref{eq:tau2}, the expected data rate in the time-based HO skipping scenario at the position $\bsym{u}=(u,0)$, $0\leq u<l$, is approximately given by
\begin{equation}
\tau'_2(u, \lambda, \beta)
=\int_{0}^{\infty}\frac{1}{1+z}\cdot\frac{\exp\Big(-\frac{\pi\lambda}{1+K_{z, \beta}^{-1}}u^{2}\Big)}{1+K_{z, \beta}}dz,
\label{eq:tau2approx}
\end{equation}
where
\begin{equation}\label{eq:kzbeta}
K_{z, \beta}=\frac{2\pi}{\beta}z^{\frac{2}{\beta}}\csc\frac{2\pi}{\beta}.
\end{equation}
\end{corollary}
\begin{IEEEproof}
Note that in \eqref{eq:cond_expect2}, the Laplace functional of PPP $\Phi$ is considered under the condition that there are no points of $\Phi$ inside the circle region $b(\mbox{\boldmath $o$}, r)$. Here, we ignore this condition and apply the Laplace functional of a homogeneous PPP over entire $\R^2$.
Then, 
\begin{align}\label{eq:cond_expect3}
     &\Exp\biggl[
       \prod_{i\in\N\setminus\{b_o\}}
         \Bigl\{
           1 + z\,\Bigl(\frac{w_{u,r,\theta}}{R_{u,i}}\Bigr)^\beta
         \Bigr\}^{-1}
     \biggm| R_{b_o}=r\biggr]
     \nonumber\\
     &\approx 
     \exp\biggl(
       -2\pi\lambda
         \int_0^\infty
           \Bigl\{
             1 + \frac{1}{z}\,
                 \Bigl(\frac{v}{w_{u,r,\theta}}\Bigr)^\beta
           \Bigr\}^{-1}\,
         v\,\dd v
     \biggr),
\end{align}
Substituting $t = z^{-1}(v/w_{u,r,\theta})^\beta$ in the integral above, we
have
\begin{align}\label{eq:cosec}
  \int_0^\infty
    \Bigl\{
      1 + \frac{1}{z}\,
          \Bigl(\frac{v}{w_{u,r,\theta}}\Bigr)^\beta
    \Bigr\}^{-1}\,
  v\,\dd v
  &= \frac{w_{u,r,\theta}^2\, z^{2/\beta}}{\beta}
     \int_0^\infty
       \frac{t^{2/\beta-1}}{1+t}\,
     \dd t
  \nonumber\\
  &= \frac{\pi\,w_{u,r,\theta}^2\, z^{2/\beta}}{\beta}\,
     \csc\frac{2\pi}{\beta}.
\end{align}
Substituting \eqref{eq:cond_expect}, \eqref{eq:cond_expect2}, \eqref{eq:cond_expect3}, and \eqref{eq:cosec} into
\eqref{eq:xi2}, using \eqref{eq:wcos}, and assuming $\sigma^2=0$, we have
\begin{align*}
&\Exp[\xi_2(u, \beta)]
\\
&\approx
\lambda\int_0^\infty\frac{\exp{(-K_{z, \beta}\,\pi\lambda u^2)}}{1+z}
\\
&\qquad\mbox{}\times\int_0^{2\pi}\!\!\!\!\int_0^\infty r
  \ee^{-\pi\lambda\big\{(1+K_{z, \beta})r^2-2K_{z, \beta}\,ur\cos\theta\big\}}
\dd r\,\dd\theta\,\dd z,
\end{align*}
where $K_{z, \beta}$ is given in \eqref{eq:kzbeta}.
Then, we can apply the following formula to the expression above:
\begin{align*}
&\int_{0}^{2\pi}\!\!\!\!\int_{0}^{\infty}r \ee^{-pr^{2}+qr\cos\theta}\dd r\,\dd\theta
\qquad\mbox{}(p>0)
\\
&\qquad\mbox{}= \int_{-\infty}^\infty\int_{-\infty}^\infty\ee^{-p(x^2+y^2)+qx}\dd x\,\dd y
\\
&\qquad\mbox{}= \ee^{q^{2}/4p}\int_{-\infty}^\infty\int_{-\infty}^\infty\ee^{-p(x^2+y^2)}\dd x\,\dd y
\\
&\qquad\mbox{}= \frac{\pi}{p}\exp{\frac{q^{2}}{4p}}\,,
\nonumber
\end{align*}
and the result follows.
\end{IEEEproof} 
\begin{remark}
In the proof of Corollary~\ref{cor:datarate2}, we approximate
the integral over $\R^2\setminus b(\mbox{\boldmath $o$}, r)$ by that over $\R^2$.
Similar approximate approaches are found in much literature (e.g., \cite{NguyBaccKofm07}).
\end{remark}

\begin{figure}[!t]
\begin{center}
\includegraphics[clip,width=.45\linewidth]{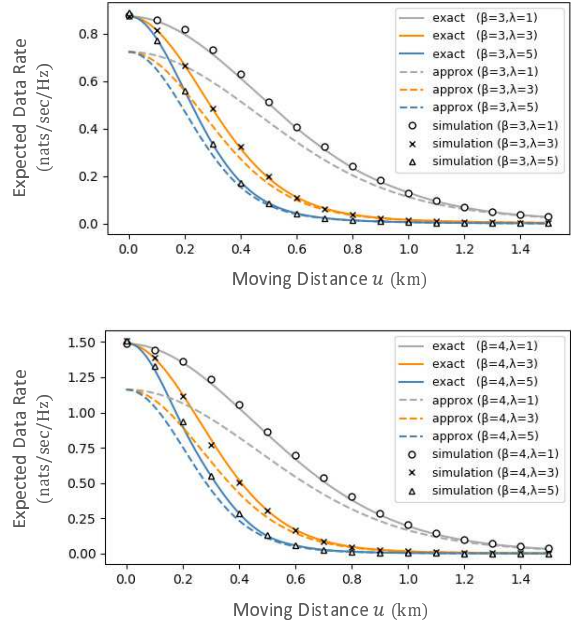}
\end{center}
\caption{Comparison among the analysis of $\tau_2(u,\lambda,\beta)$ in \eqref{eq:tau2}, the approximate analysis of $\tau'_2(u,\lambda,\beta)$ in \eqref{eq:tau2approx}, and the simulated result of $\tau_2(u,\lambda,\beta)$ using the mean of 10,000~independent samples through simulation. In the graph legends, exact and approx correspond to  $\tau_2(u,\lambda,\beta)$ and $\tau'_2(u,\lambda,\beta)$, respectively.}
\label{fig:xi2}
\end{figure}

Fig.~\ref{fig:xi2} shows a comparison among the numerically
computed $\tau_2(u,\lambda,\beta)$ in \eqref{eq:tau2}, $\tau'_2(u,\lambda,\beta)$ in \eqref{eq:tau2approx}, and the mean of
10,000~independent samples of $\xi_2(u,\beta)$ in~\eqref{eq:xi20}
computed by Monte Carlo simulation.
We see that the analytical result of $\tau_2(u,\lambda,\beta)$ surely corresponds to the simulation.
On the other hand, the analytical result of $\tau'_2(u,\lambda,\beta)$ has some gaps from the result of $\tau_2(u,\lambda,\beta)$,
particularly when the moving distance of the UE is small.
However, these gaps tend to be small as the moving distance
becomes large.
This is because the distance from the UE to the points of $\Phi$ in $b(\mbox{\boldmath $o$}, r)$ becomes large as the moving distance $u$ increases, and therefore the effect of the approximation becomes weak.
Furthermore, it can be seen that the expected data rate is smaller when
the BS density~$\lambda$ is larger or the path loss
exponent~$\beta$ is smaller, even for the same moving distance~$u$.
This is because the interference power increases
as $\lambda$ increases or $\beta$ decreases.

Corollary~\ref{cor:datarate2} provides a simple form for calculating the data rate in Scenario~2; however, as shown in Fig.~\ref{fig:xi2}, the introduced approximation causes some errors, particularly when a UE is close to the starting point (i.e., $u=0$). Nevertheless, since $\tau_2(0, \lambda, \beta) = \tau_1(\lambda, \beta)$ can be calculated exactly by Theorem~\ref{prp:datarate1}, we can refine the approximate formula by using $\tau_2(0, \lambda, \beta)$. 
%Nevertheless, as we know $\tau_2(0, \lambda, \beta)=\tau_1(\lambda, \beta)$, the exact expression of the data rate can be derived by Theorem~\ref{prp:datarate1} when  $u=0$, so that the error seen in Corollary~\ref{cor:datarate2} can be compensated by using $\tau_1(\lambda, \beta)$ of Theorem~\ref{prp:datarate1}. 
Accordingly, we propose a refined version of Corollary~\ref{cor:datarate2} as follows.
\begin{corollary}
$\tau_2(u, \lambda, \beta)$ in \eqref{eq:tau2}, the expected data rate in the time-based HO skipping scenario at the position $\bsym{u}=(u,0)$, $0\leq u<l$, is approximately given by
\begin{equation}\label{eq:tau2approxex}
\tau''_2(u, \lambda, \beta, \varepsilon)
:=\ee^{-\varepsilon u}\tau_1(\lambda, \beta) + (1-\ee^{-\varepsilon u})\tau'_2(u, \lambda, \beta),
\end{equation}
where $\tau_1(\lambda, \beta)$, and $\tau'_2(u, \lambda, \beta)$ are in \eqref{eq:tau1}, and \eqref{eq:tau2approx}, respectively. Moreover, $\varepsilon$ is a fitting parameter.
\end{corollary}
\begin{remark}
We note that the form in \eqref{eq:tau2approxex} takes its value between $\tau_1(\lambda, \beta)$ and $\tau'_2(u, \lambda, \beta)$, and the balance of the weight between the two metrics is gradually inclined from the former to the latter as $u$ becomes large. The degree of the incline is adjusted by the fitting parameter $\varepsilon$. This refines the error in \eqref{eq:tau2approx} because $\tau''_2(0, \lambda, \beta, \varepsilon)=\tau_1(\lambda, \beta)$, and $\tau''_2(u, \lambda, \beta, \varepsilon)\approx\tau'_2(u, \lambda, \beta)$ for sufficiently large $u$, where we recall that $\tau'_2(u, \lambda, \beta)$ tends to be exact as $u$ becomes large.
\end{remark}

\begin{figure}[!t]
\begin{center}
\includegraphics[clip,width=.4\linewidth]{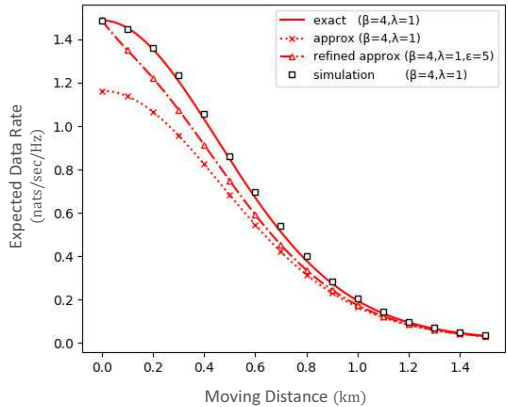}
\end{center}
\caption{Comparison among $\tau_2(u, \lambda, \beta)$, $\tau'_2(u, \lambda, \beta)$, $\tau''_2(u, \lambda, \beta)$ (the three results of analysis of the expected data rate) and the mean of 10,000~independent samples
  of $\xi_2(u,\beta)$ through simulation. In the graph legends, exact, approx, and refined approx correspond to $\tau_2(u,\lambda,\beta)$ in \eqref{eq:tau2}, $\tau'_2(u,\lambda,\beta)$ in \eqref{eq:tau2approx}, and $\tau''_2(u,\lambda,\beta,\varepsilon)$ in \eqref{eq:tau2approxex},  respectively. The fitting parameter $\varepsilon$ is set as $\varepsilon=5$.}
\label{fig:drcomp2}
\end{figure}

Fig.~\ref{fig:drcomp2} shows numerically computed values of the expected data rate at $(u, 0)$ based on three different values of $\tau_2(u, \lambda, \beta)$ in \eqref{eq:tau2}, $\tau'_2(u, \lambda, \beta)$ in \eqref{eq:tau2approx}, and $\tau''_2(u, \lambda, \beta, \varepsilon)$ in \eqref{eq:tau2approxex}, compared with the simulation result.
We set the fitting parameter as $\varepsilon=5$. It can be seen that the curve of $\tau''_2$ is closer to the exact result $\tau_2$ than the approximation result $\tau'_2$. 
%We note that the curve of $\tau''_2$ becomes closer to that of $\tau_2$, the exact result, as the parameter $\varepsilon$ is more adjusted.  

As described above, $\tau''_2(u, \lambda, \beta, \varepsilon)$ in \eqref{eq:tau2approxex} provides well-accurate result of the data rate in Scenario~2 with less computational complexity, although the parameter $\varepsilon$ needs fitting for $\lambda$ and $\beta$. In contrast, such fitting is not needed for $\tau_2(u, \lambda, \beta)$ in \eqref{eq:tau2} and $\tau'_2(u, \lambda, \beta)$ in \eqref{eq:tau2approx}, whereas the former has serious computational complexity, and the latter has some errors from the exact result.

\section{Evaluation of Transmission Performance}\label{sec:evaluation}

In this section, we study the transmission performance for a moving UE on the basis of the analysis in the preceding section. We first compare the transmission performance between Scenarios~1 and 2. Next, we consider optimization of the performance in Scenario~2.
Our purpose in this section is to see whether the time-based HO skipping scenario outperforms the non-skipping scenario, and to find an optimal skipping time that improves the transmission performance in the time-based HO skipping scenario. 
%the performance is improved by adjusting the skipping time in the time-based HO skipping scenario.

\subsection{Performance Comparison of Non-skipping and Time-based HO Skipping Scenarios}

We here compare the performance of Scenario~1 and~2 by using the evaluation functions $Q_1(s, \lambda,\beta,\mathcal{C})$ and $Q_2(s, \lambda, \beta, \mathcal{C})$ given in \eqref{eq:tpfunc}. As for $Q_1(s, \lambda,\beta,\mathcal{C})$, $\Exp[\mathcal{N}_1(L_1,\lambda)] = 4\sqrt{\lambda}\,\Exp[L_1]/\pi$ holds by \eqref{eq:handover1}, so that $Q_1(s, \lambda,\beta,\mathcal{C})$ depends on $L_1$ only through its mean. Whereas, $Q_2(s,\lambda,\beta,\mathcal{C})$ depends on the distribution of $L_1$. We should also note that since $L_n$, $n=1,2,\cdots$, are i.i.d.\ and the system model is stationary and isotropic, $Q_m(s, \lambda,\beta,\mathcal{C})$, $m=1,2$, represents the transmission performance in a movement period averaged over the entire trajectory of a moving UE. 

\begin{figure}[!t]
\begin{center}
\includegraphics[clip,width=.45\linewidth]{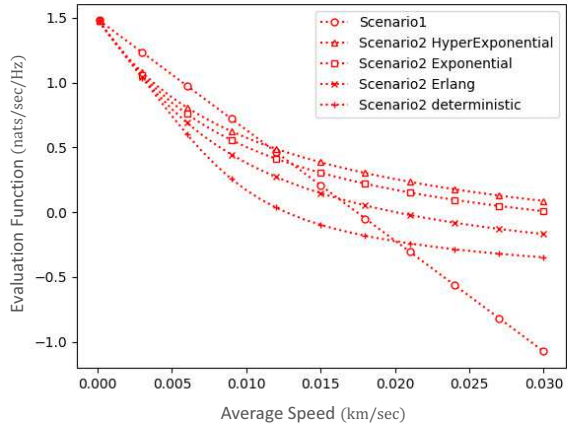}
\end{center}
\caption{Comparison of $Q_1$ and $Q_2$, where the case of deterministic~$L_1=l$ and several cases of randomly distributed~$L_1$ with mean~$l$ are examined. $\tau''_2(u,\lambda,\beta)$ (the approximate result in \eqref{eq:tau2approxex}) is applied to the expected data rate of $Q_2$. The fitting parameter of $\tau''_2$ is $\varepsilon=10$, and the other parameters are set as $s=50$ (sec), $\lambda=5$ (units/km$^2$), and $(\beta, \mathcal{C}, \sigma^2) = (4, 30, 0)$. As for the Erlang distribution, the shape parameter is set as $k=2$. As for the hyper-exponential distribution, two exponential distributions are mixed, and their intensities $\lambda_1, \lambda_2$ obey the law $\lambda_1=3\lambda_2$. The mixture probability is $p'=1/2$.}
\label{fig:expectedQ}
\end{figure}

Fig.~\ref{fig:expectedQ} shows a comparison of the evaluation function among Scenario~1 and Scenario~2, where $Q_1(s, \lambda,\beta,\mathcal{C})$ is computed for Scenario~1, and $Q_2(s, \lambda, \beta, \mathcal{C})$ is computed for Scenario~2 with $\tau_2(u,\lambda,\beta)$ replaced by $\tau''_2(u,\lambda,\beta,\varepsilon)$ in \eqref{eq:tau2approxex}.
In Fig.~\ref{fig:expectedQ}, $L_1$ follows four different distributions; exponential distribution, Erlang distribution, hyper-exponential distribution, and deterministic case, any of which have the same mean value $l=\Exp[L_1]$. There is only one curve exhibited for $Q_1$ because $Q_1$ depends on $L_1$ only through its mean, and thus all the four cases are the same for $Q_1$. The horizontal axis in the figure denotes $l=\Exp[L_1]$.
From Fig.~\ref{fig:expectedQ}, we conclude that Scenario~1 is better when the average speed of the moving UE is sufficiently slow, whereas Scenario~2 is preferred when the UE moves fast. This is because the cost of performing an HO becomes larger as the UE moves faster, and therefore the performance of the HO skipping scenario outperforms the non-skipping scenario.
Moreover, the figure shows that the distribution of the moving speed of the UE has a large impact on the evaluation function in Scenario~2 even if the average speed is the same. In particular, the evaluation function in Scenario~2 becomes larger as $L_1$ becomes larger in the convex order (see, e.g. \cite{ShakShan07}), that is, the expectation of the evaluation function becomes larger as the distribution of $L_1$ exhibits stronger randomness. 
%We expect the reason as follows; as the variance of the moving speed becomes large in the HO skipping scenario, the situation that the UE is located closely to its connected BS is likely to occur even when the average speed is the same, which dominates the expectation  Since the HO probability and the path-loss effect to the data rate has an exponential 
%This leads to larger expected received data and smaller HO probability.
%because the exponential path-loss effect is assumed in our model.,  the path-loss affects the data rate exponentially , which leads to larger expected recieved data and smaller HO probability.

\subsection{Optimizing Performance in Time-based HO Skipping Scenario}

In this section, we further evaluate the transmission performance in Scenario~2 to characterize the time-based HO skipping. 
Our main purpose in this section is to see whether there is an optimal skipping time that maximizes the evaluation function $Q_2(s, \lambda, \beta, \mathcal{C})$ given in \eqref{eq:tpfunc} with respect to $s$. 
Here, we consider a fixed moving distance in a movement period as in Section~\ref{sec:analysis}, so that the speed of a UE in a movement period is equal to a constant $v=l/s$.
To emphasize the dependence on the moving speed $v$, we define $Q_{v, \lambda, \beta, \mathcal{C}}(s):=Q_2(s, \lambda, \beta, \mathcal{C})$, which we regard as a function of $s$. The evaluation function $Q_{v, \lambda, \beta, \mathcal{C}}(s)$ is expressed as
\begin{align}
Q_{v, \lambda, \beta, \mathcal{C}}(s)
&=\frac{1}{s}\big\{D_{2}(s, l, \lambda, \beta)-\mathcal{C}\mathcal{N}_{2}(l, \lambda)\big\}
\nonumber\\
&=\frac{1}{s}\sum_{t=0}^{s-1}\tau_2(v t, \lambda, \beta)-\frac{\mathcal{C}}{s}\mathcal{N}_2(sv, \lambda),
\label{eq:Q_discrete}
\end{align}
where \eqref{eq:D20} is applied in the second equality, and $\tau_2(sw, \lambda, \beta)$ and $\mathcal{N}_2(sv, \lambda)$ are respectively given in \eqref{eq:tau2} and \eqref{eq:handover2}. Although the parameter $s$ in the function above is a positive integer, we here modify the function so as to make $s$ continuous as follows;
we regard the sum on the right-hand side of \eqref{eq:Q_discrete} as a
Riemann sum and approximate it by the corresponding integral, that is,
\begin{equation}\label{eq:Riemann_approx}
  \sum_{t=0}^{s-1}
      \tau_2\Bigl(\frac{l t}{s}, \lambda, \beta\Bigr)
  \approx
    \frac{s}{l}
    \int_0^l \tau_2(u, \lambda, \beta)\,\dd u.
\end{equation}  
In addition, we replace $\tau_2(sw, \lambda, \beta)$ in \eqref{eq:Q_discrete} with $\tau'_2(sw, \lambda, \beta)$ in \eqref{eq:tau2approx} for the sake of simplicity. 
%For the sake of simplicity of its form, we consider replacing $\tau_2(sw, \lambda, \beta)$ with $\tau_2'(sw, \lambda, \beta)$ given in \eqref{eq:tau2approx}. 
Then, using $l=sv$, we obtain an approximate continuous function of $Q_{v, \lambda, \beta, \mathcal{C}}(s)$  as follows:
%Then, substituting \eqref{eq:Riemann_approx} to \eqref{eq:Q_discrete} and using $l=sv$, we derive a continuous form of the function $\widetilde{Q}(s)\approx Q_{v, \lambda, \beta, \mathcal{C}}(s)$ as
\begin{equation}
\label{eq:Q_continuous}
Q_{v, \lambda, \beta, \mathcal{C}}(s)
\approx
\widetilde{Q}(s)
:=
\frac{1}{v}\int_0^{v}\tau_2'(sw, \lambda, \beta)\dd w-\frac{\mathcal{C}}{s}\mathcal{N}_2(sv, \lambda).
\end{equation}

\begin{figure}[!t]
\begin{center}
%\centering
\includegraphics[width =.47\linewidth, clip]{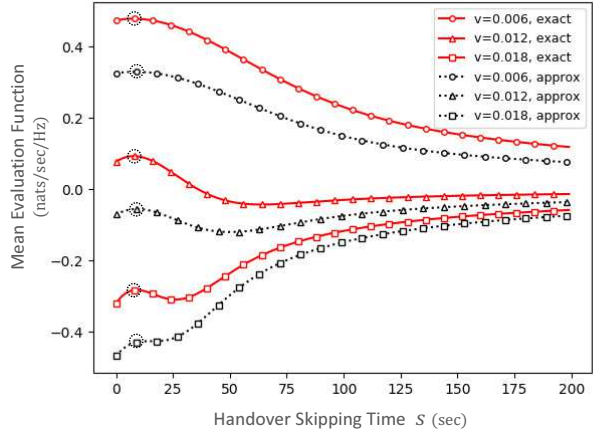}
\vspace{-1.30mm}
\caption{Numerical samples of $Q_{v\!,\lambda,\beta\!,\mathcal{C}}(s)$ and $\widetilde{Q}(s)$, which are plotted in the red solid lines and the black dotted lines, respectively. The numerical samples are computed with several cases of $v$ and fixed $\lambda, \beta, \mathcal{C}$ as $\lambda=3$ (units/km$^2$), and $(\beta, \mathcal{C}) = (3, 30)$.
\label{fig:Qsgraph2-4}
}
\end{center}
\vspace{-3.80mm}
\end{figure}

As demonstrated in Fig.~\ref{fig:drcomp2}, the approximation $\tau_2'(sw, \lambda, \beta)$ contains some errors. Thus, we here discuss the approximation error of $\widetilde{Q}(s)$
compared to $Q_{v, \lambda, \beta, \mathcal{C}}(s)$. 
Fig.~\ref{fig:Qsgraph2-4} shows comparison results of the evaluation functions $Q_{v, \lambda, \beta, \mathcal{C}}(s)$ and $\widetilde{Q}(s)$, which are numerically computed with different $v$ and fixed $\lambda, \beta$, and $\mathcal{C}$.
From the figure, we see that both the evaluation functions $Q_{v, \lambda, \beta, \mathcal{C}}(s)$ and $\widetilde{Q}(s)$ have similar shapes, although there are some errors between those two functions. In particular, both $Q_{v, \lambda, \beta, \mathcal{C}}(s)$ and $\widetilde{Q}(s)$ exhibit a local maximum with respect to the skipping time $s$. 
In addition, both of the local optimal skipping times for $Q_{v, \lambda, \beta, \mathcal{C}}(s)$ and $\widetilde{Q}(s)$ take similar values. 
Also, the local optimal skipping time is almost invariant with respect to $v$ exhibited in the figure.
Furthermore, since $Q_{v, \lambda, \beta, \mathcal{C}}(s)$ and $\widetilde{Q}(s)$ converge to zero as $s$ increases, the local maximum can be regarded as the global maximum when $v$ is sufficiently small. Therefore, the results indicate that there is an optimal skipping time that improves the evaluation function of transmission performance of a moving UE.
Because both the exact and approximate evaluation functions have similar optimal skipping time, we henceforth evaluate the optimal skipping time $s^\ast$ that maximizes $\widetilde{Q}(s)$ instead of $Q_{v, \lambda, \beta, \mathcal{C}}(s)$.

\begin{figure}[!t]
\begin{center}
%\centering
\includegraphics[width =.47\linewidth, clip]{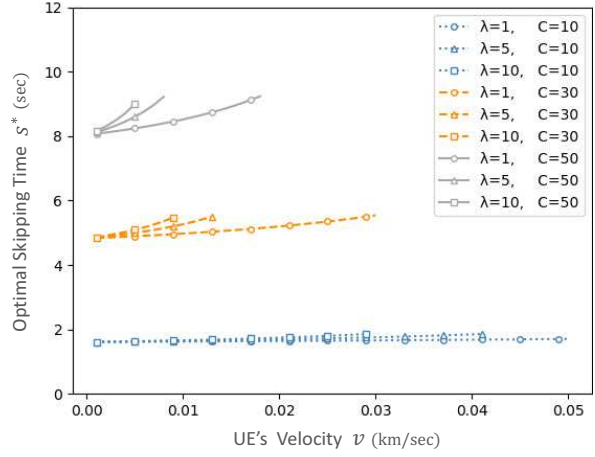}
\vspace{-1.30mm}
\caption{Numerically computed results of $s^*$ with several patterns of the parameters $\lambda$ and $\mathcal{C}$, and with fixed parameter $\beta$ as $\beta=4$. 
\label{fig:opts1v}
}
\end{center}
\vspace{-3.80mm}
\end{figure}

\begin{figure}[!t]
\begin{center}
%\centering
\includegraphics[width =.47\linewidth, clip]{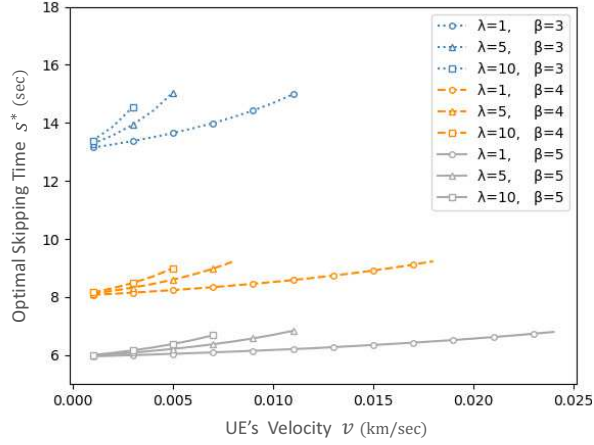}
\vspace{-1.30mm}
\caption{Numerically computed results of $s^*$ with several cases of the parameters $\lambda$ and $\beta$, and with fixed parameter $\mathcal{C}$ as $\mathcal{C}=50$.
\label{fig:opts2lam}
}
\end{center}
\vspace{-3.80mm}
\end{figure}

\begin{figure}[!t]
\begin{center}
%\centering
\includegraphics[width =.47\linewidth, clip]{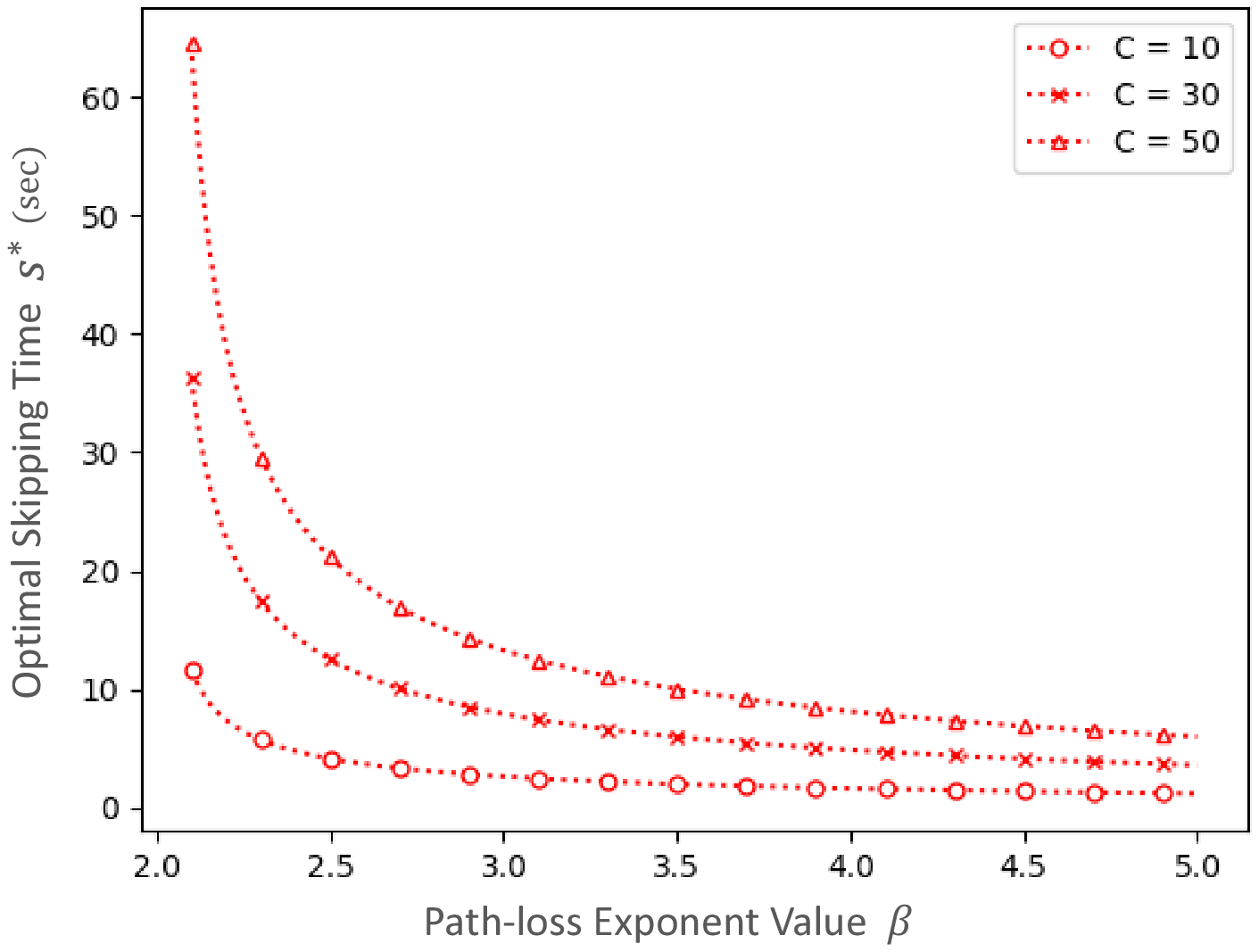}
\vspace{-1.30mm}
\caption{Numerically computed results of $s^*$ with three cases of the parameter $\mathcal{C}$, and with fixed parameters $v$, $\lambda$ as $v=0.001$~(km/sec), $\lambda=5$~(units/km$^2$).
\label{fig:opts3beta}
}
\end{center}
\vspace{-3.80mm}
\end{figure}

\begin{figure}[!t]
\begin{center}
%\centering
\includegraphics[width =.47\linewidth, clip]{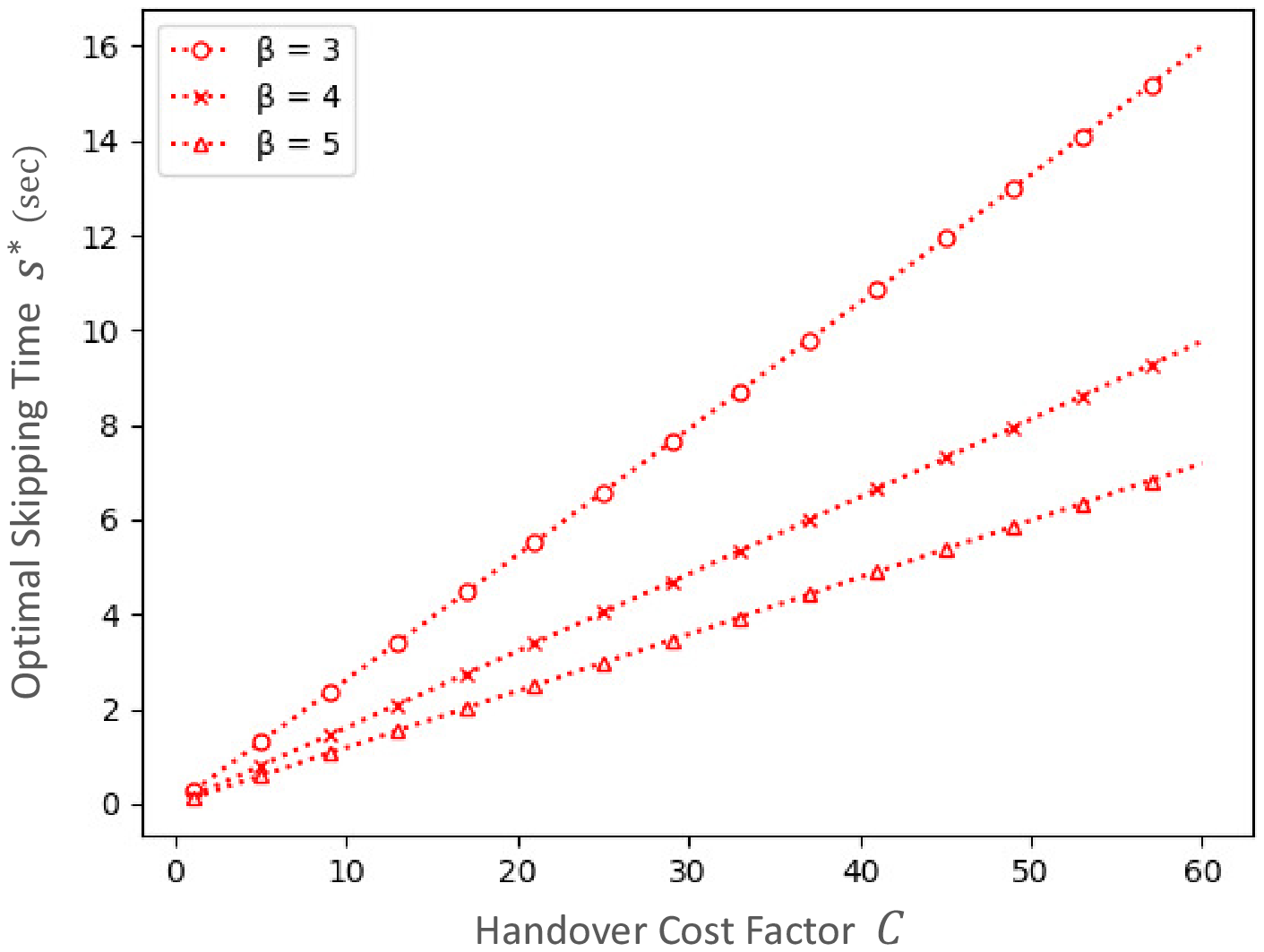}
\vspace{-1.30mm}
\caption{Numerically computed results of $s^*$ with three cases of the parameter $\beta$, and with fixed parameters $v$, $\lambda$ as $v=0.001$~(km/sec), $\lambda=5$~(units/km$^2$).
\label{fig:opts4C}
}
\end{center}
\vspace{-3.80mm}
\end{figure}

We now investigate $s^*$ by evaluating the impact of the other system parameters. We numerically compute $s^*$ and compare the results among several patterns for the values of $v, \lambda, \beta$, and $\mathcal{C}$. Figs.~\ref{fig:opts1v},~\ref{fig:opts2lam},~\ref{fig:opts3beta}, and~\ref{fig:opts4C} show numerical results of the value of $s^*$. The horizontal axes in these figures denote $v$ in Fig.~\ref{fig:opts1v},~\ref{fig:opts2lam}, $\beta$ in Fig.~\ref{fig:opts3beta}, and $\mathcal{C}$ in Fig.~\ref{fig:opts4C}, respectively, and the remaining parameters are fixed for each curve. We note that the curves are terminated when $s^*$ does not exist on $\widetilde{Q}(s)$, which occurs when any of the parameters $v, \lambda, \beta$, and $\mathcal{C}$ exceed a certain threshold; $s^*$ does not exist when $v, \lambda$ or $\mathcal{C}$ is sufficiently large, or $\beta$ is sufficiently small. 
From the figures, we see that $s^*$ changes drastically with respect to $\beta$ or $\mathcal{C}$, whereas $s^*$ changes slightly with respect to $v$ or $\lambda$.
The reason for this is that the data rate and the HO rate, the two factors constituting the evaluation function $\widetilde{Q}(s)$ in \eqref{eq:Q_continuous}, are affected by similar orders with respect to $v$ and $\lambda$, thus the two factors cancel each other.
Moreover, from Fig.~\ref{fig:opts3beta}, we see that $s^*$ becomes small as $\beta$ increases. 
As we see in Fig.~\ref{fig:xi2}, the data rate $\tau_2'(sw, \lambda, \beta)$ decreases more rapidly with respect to the skipping time $s$ when $\beta$ is larger. Therefore, smaller $s$ provides better results when $\beta$ is larger. Furthermore, we see from Fig.~\ref{fig:opts4C} that $s^*$ becomes large as $\mathcal{C}$ increases. This is intuitively because increasing $\mathcal{C}$ makes the cost of HOs larger, and therefore longer skipping time is preferred to improve the performance. 
Analytical studies about these behaviors of $s^*$ are given in the next section.  


\subsection{Analytical Expressions of the Optimal Skipping Time}

As shown in the numerical results in the previous section, the evaluation function in the time-based HO skipping scenario can be maximized with respect to the skipping time $s$. An interesting feature in these results is that the skipping time that gives the maximum is almost invariant with respect to the moving speed $v$. On the basis of this observation, we derive the theoretical expressions of the optimal skipping time $s^*$ approximately under the condition that the moving speed $v$ is asymptotically close to zero. Then, we apply them to the general cases of $v$. 
The approximate expression of $s^*$ is given as follows.
\begin{proposition}
If $v\approx0$, then the value $s^*$ that maximizes $\widetilde{Q}(s)$ in $\eqref{eq:Q_continuous}$ is expressed as follows:
\begin{equation}
s^*\approx\frac{15-\pi^2}{4\pi^2}\mathcal{C}\times\Big\{\int_{0}^{\infty}\frac{K_{z, \beta}}{(1+z)(1+K_{z, \beta})^2}\dd z\Big\}^{-1},
\label{eq:optsEST}
\end{equation}
where $K_{z, \beta}$ is given in \eqref{eq:kzbeta}.
\end{proposition}
\begin{IEEEproof}
To obtain the first derivative of $\widetilde{Q}(s)$ with respect to $s$, we consider the Taylor expansion around $v=0$, and ignore terms for which the order of $v$ is greater than 2. Then, the first term of $\widetilde{Q}(s)$ yields
\begin{align}
&\!\!\!\frac{1}{v}\cdot\frac{\dd}{\dd s}\int_0^{v}\tau_2(sw, \lambda, \beta)\;\dd w
\nonumber\\
&=\frac{1}{v}\cdot\frac{\dd}{\dd s}\int_{0}^{v}\!\!\!\!\int_{0}^{\infty}\frac{1-\frac{\pi\lambda}{1+K_{z, \beta}^{-1}}s^2w^2}{(1+z)(1+K_{z, \beta})}\dd z\dd w
+o(v^2)\;\;as\;\;v\rightarrow0
\nonumber\\
&\approx-\frac{2}{3}sv^2\pi\lambda\int_0^\infty\frac{K_{z, \beta}}{(1+z)(1+K_{z, \beta})^2}\dd z.
\label{eq:derivative1}
\end{align}
Regarding the second term of $\widetilde{Q}(s)$ in \eqref{eq:Q_continuous}, we consider the asymptotical form of $\mathcal{N}_2(sv, \lambda)$ in \eqref{eq:handover2} under the condition that $v\approx0$. Since $v\ll r$ yields $\sin^{-1}\big(\frac{sv\sin\theta}{w_{sv,r,\theta}}\big)\approx\sin^{-1}\big(\frac{sv\sin\theta}{r}\big)$, $B(sv, r, \theta)$ in \eqref{eq:regionB} is expressed as follows
\begin{align*}
&B(sv, r, \theta)
\\
&\approx{w_{sv, r, \theta}}^2\Big[\theta+\sin^{-1}\Big(\frac{sv\sin\theta}{r}\Big)\Big]-r^2\theta+svr\sin\theta
\\
&={w_{sv, r, \theta}}^2\Big[\theta+\frac{sv\sin\theta}{r}\Big]-r^2\theta+svr\sin\theta+o(v^2)
\\
&=-2svr(\theta\cos\theta-\sin\theta)+s^2v^2(\theta-\sin2\theta)+o(v^2)\;\;as\;\;v\rightarrow0,
%\\
%&\qquad\qquad\qquad\qquad\qquad\quad+\mathcal{O}(v^3)\;\;as\;\;v\rightarrow0,
\end{align*}
where the Maclaurin expansion of $\sin^{-1}\big(\frac{sv\sin\theta}{r}\big)$ is applied in the first equality, and \eqref{eq:wcos} is used in the second equality. Substituting the above expression into \eqref{eq:handover2} yields
\begin{align*}
&\mathcal{N}_2(sv, \lambda)
\\
&\approx
 1 -\! 2\lambda \!\!
 	\int^{\pi}_{0} \!\!\!\!
 		\int^{\infty}_{0} \!\!\!
 			r\ee^{-\lambda\pi r^{2}}
\\
&\quad\times\ee^{2\lambda svr(\theta\cos\theta-\sin\theta)}\cdot\ee^{-\lambda s^2v^2(\theta-\sin2\theta)}
 		\dd r
	\dd \theta
	+o(v^2)
%\\
%&\qquad\qquad\qquad\qquad\qquad\quad\qquad\qquad+\mathcal{O}(v^3)
\\
&=
 -\, 2\lambda \!
 	\int^{\pi}_{0} \!\!\!\!
 		\int^{\infty}_{0} \!\!
 			r\ee^{-\lambda\pi r^{2}}\Big[2\lambda svr(\theta\cos\theta-\sin\theta)
\\
&\quad+\lambda s^2v^2\big\{2\lambda r^2(\theta\cos\theta-\sin\theta)^2-(\theta-\sin2\theta)\big\}\Big]
 		\dd r
	\dd \theta
	+o(v^2)
%\\
%&\qquad\qquad\qquad\qquad\qquad\quad\qquad\qquad+\mathcal{O}(v^3)
\\
&=\frac{4\sqrt{\lambda}}{\pi}sv-\frac{15-\pi^2}{6\pi}\lambda s^2v^2+o(v^2)\;\;as\;\;v\rightarrow0,
\end{align*}
where the Maclaurin expansion of the exponential form is applied in the first equality.
Therefore, we have
\begin{equation}\label{eq:derivative2}
\frac{\dd}{\dd s}\cdot\frac{\mathcal{C}}{s}\mathcal{N}_2(sv, \lambda)
\approx-\frac{15-\pi^2}{6\pi}\mathcal{C}\lambda v^2.
\end{equation}
From \eqref{eq:derivative1} and \eqref{eq:derivative2}, we derive the asymptotical form of the first-order derivative of $\widetilde{Q}(s)$ as
\begin{align}
&\!\frac{\dd}{\dd s}\widetilde{Q}(s)
\nonumber\\
&\approx-\frac{2}{3}sv^2\pi\lambda\int_{0}^{\infty}\!\!\frac{K_{z, \beta}}{(1+z)(1+K_{z, \beta})^{2}}dz
+\frac{15-\pi^2}{6\pi}\mathcal{C}v^2\lambda.
\nonumber
\end{align}
It follows from the expression above that the approximate form of the function $\widetilde{Q}(s)$ is concave and has only one peak because $\frac{\dd}{\dd s}\widetilde{Q}(s)$ is a linear function with respect to $s$, and $\frac{\dd}{\dd s}\widetilde{Q}(s)>0$ when $s$ is sufficiently small and $\frac{\dd}{\dd s}\widetilde{Q}(s)<0$ when $s$ is sufficiently large. Thus, the extreme point of $\widetilde{Q}(s)$ corresponds to the maximum with respect to $s$.

Finally, solving $\frac{\dd}{\dd s}\widetilde{Q}(s)=0$ for $s$ completes the proof.
\end{IEEEproof}

\begin{figure}[!t]
\begin{center}
%\centering
\includegraphics[width =.47\linewidth, clip]{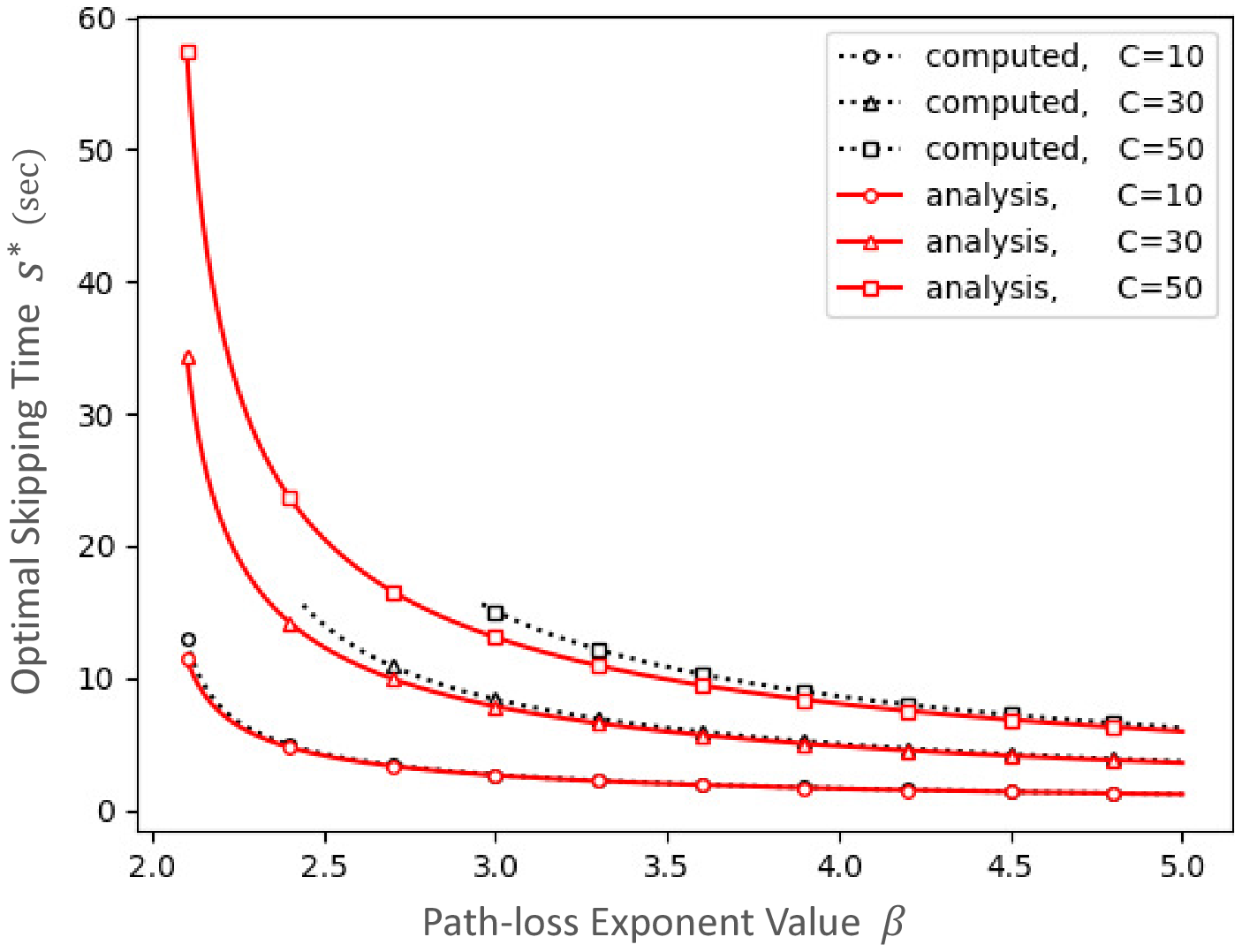}
\vspace{-1.30mm}
\caption{Comparison results among the analysis in \eqref{eq:optsEST} and $s^*$ obtained by numerical searching computation of $\widetilde{Q}$. The exhibited curves refer to three cases of the parameter $\mathcal{C}$, and fixed parameters $v$, $\lambda$ as $v=0.005$ (km/sec), $\lambda=5$ (units/km$^2$).
\label{fig:opts1estimate}
}
\end{center}
\vspace{-3.80mm}
\end{figure}
\begin{figure}[!t]
\begin{center}
%\centering
\includegraphics[width =.47\linewidth, clip]{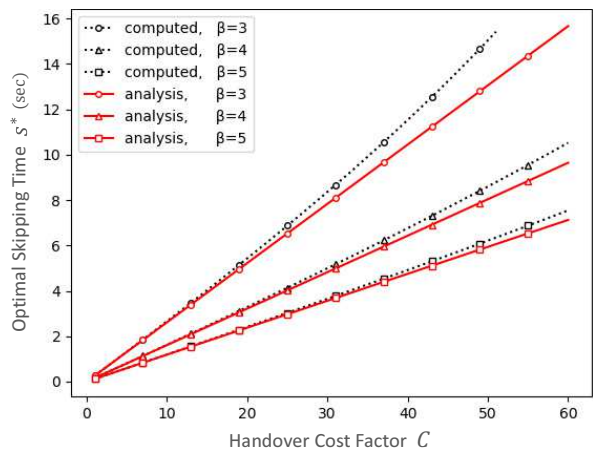}
\vspace{-1.30mm}
\caption{Comparison results among the analysis in \eqref{eq:optsEST} and $s^*$ obtained by numerical searching computation of $\widetilde{Q}$. The exhibited curves refer to three cases of the parameter $\beta$, and fixed parameters $v$, $\lambda$ as $v=0.005$ (km/sec), $\lambda=5$ (units/km$^2$).
\label{fig:opts2estimate}
}
\end{center}
\vspace{-3.80mm}
\end{figure}

\begin{remark}
We here discuss the accuracy of the approximate expression in \eqref{eq:optsEST}. The expression indicates that the optimal skipping time $s^*$ depends only on $\beta$ and $\mathcal{C}$, not on $\lambda$. Figs.~\ref{fig:opts1v} and~\ref{fig:opts2lam} demonstrate that $s^*$ slightly changes with respect to $\lambda$ even if $v$ is large. Therefore, the approximate expression \eqref{eq:optsEST} is expected to have sufficient accuracy in terms of the parameter $\lambda$ when we apply \eqref{eq:optsEST} for general $v$.
We thus examine the accuracy of the expression in terms of $\beta$ and $\mathcal{C}$. Figs.~\ref{fig:opts1estimate} and~\ref{fig:opts2estimate} show a comparison between the approximate expression of $s^*$ in \eqref{eq:optsEST} and the exact value of $s^*$ obtained by numerical computation. The horizontal axes in Figs.~\ref{fig:opts1estimate} and~\ref{fig:opts2estimate} denote $\beta$, and $\mathcal{C}$ respectively.
We conclude that the expression in \eqref{eq:optsEST} fits to the numerical result with small errors, even when $v$ is not close to zero. Thus, it appears that we can adopt \eqref{eq:optsEST} as an approximated result of the expression of $s^*$ for general $v, \lambda, \beta$, and $\mathcal{C}$.
\end{remark}
\begin{remark}
The expression in \eqref{eq:optsEST} particularly implies that the optimal skipping time $s^*$ increases linearly as $\mathcal{C}$ increases. This is due to the form of $\widetilde{Q}(s)$; the evaluation function $\widetilde{Q}(s)$ given in our model is an increasing linear function with respect to $\mathcal{C}$. 
\end{remark}

\section{Conclusion}\label{sec:conclusion}

In this paper, we proposed a novel HO skipping scheme, {\it time-based} HO skipping, and provided a random walk mobility model for a moving UE that performs the time-based HO skipping. Under the model of single-tier downlink cellular networks where BSs are allocated according to homogeneous PPP, we derived theoretical expressions of the two metrics: the expected data rate and the HO rate for a moving UE that performs the time-based HO skipping. On the basis of these theoretical results, we studied transmission performance of a moving UE using the two metrics, under the model of the time-based HO skipping.
From our analysis, we found that the transmission performance in the time-based HO skipping scenario can outperform the non-skipping scenario when the moving speed of a UE is sufficiently large. Moreover, we found that there is a local maximum of the transmission performance with respect to the skipping time of a UE. We also investigated to derive a theoretical expression of the skipping time that gives the maximum.

For further study, our mobility model assumed that a UE moves along a straight line during a movement period, that is, a UE cannot moves along a curve in our model. Therefore, it would be expected to extend our analysis to more general trajectory model of a moving UE.

\end{document}